# Improving Power Flow Robustness via Circuit Simulation Methods

Amritanshu Pandey[1], Marko Jereminov[1], Gabriela Hug[2,1], Larry Pileggi[1]

[1]Dept. of Electrical and Computer Engineering
Carnegie Mellon University
Pittsburgh, PA

[2]Power Systems Laboratory
ETH Zurich
Switzerland

*Abstract-* Recent advances in power system simulation have included the use of complex rectangular current and voltage (I-V) variables for solving the power flow and three-phase power flow problems. This formulation has demonstrated superior convergence properties over conventional polar coordinate based formulations for three-phase power flow, but has failed to replicate the same advantages for power flow in general due to convergence issues with systems containing PV buses. In this paper, we demonstrate how circuit simulation techniques can provide robust convergence for any complex I-V formulation that is derived from our split equivalent circuit representation. Application to power grid test systems with up to $10^4$ buses demonstrates consistent global convergence to the correct physical solution from arbitrary initial conditions.

## I. Introduction

The industry standard for solving the power flow problem is the polar coordinate based "PQV" formulation with real power (P), reactive power (Q), voltage (V) and angle (δ) as the state variables. This formulation is most applicable when the power grid is modeled exclusively using constant power load and generator models (PQ and PV buses). It is not, however, the optimal formulation when measurement data or physics based load models are incorporated, since these components are best modeled using the voltages and currents that are the natural state variables of the system.

Dommel first proposed the use of complex rectangular I-V variables for the power flow problem formulation in the late 1960s [1] when he proposed to model the PQ buses in the system as current injections. The proposed formulation was not widely accepted by the industry for the power flow problem due to the inability to model PV buses satisfactorily [2]. Later in 1999, a solution method was proposed formulating the governing equations for PV buses using I-V variables [3]. This approach was termed the current injection method (TCIM). In this approach, each PV bus in the system is represented by two nonlinear equations augmented with a new dependent variable of reactive power Q. Even though the proposed formulation demonstrated superior performance over conventional power flow for well-conditioned systems, it exhibited convergence issues for ill-conditioned heavily loaded systems [4]. The same authors proposed in 2004 a formulation for the PV buses (i.e. similar to the one discussed in [1]) to overcome these limitations resulting in a numerically more robust approach than the previous formulation. This improved formulation exhibited the same convergence characteristics as the conventional quadratic "PQV" formulation based on polar coordinates.

In spite of these advancements, the use of rectangular I-V formulation has been mostly restricted to solving the three-phase power distribution problem for radial distribution systems. Application of rectangular I-V formulation to a typical transmission system containing a large number of PV buses is known to cause a variety of problems [5]-[7]. With I-V formulation, each PV bus and PQ bus connected to a power electronic device augments the solution space by adding an additional unspecified Q variable. It is difficult to predict the best initial guess for these unspecified Q variables, which has been demonstrated to cause convergence issues. Furthermore, the I-V formulation is known to introduce matrix singularity conditions that result in additional convergence issues.

In this paper, we describe a robust power flow simulation approach based on the use of the natural state variables (voltages and currents). We demonstrate that this approach can facilitate incorporation of measurement-data based load models directly into the system model [8]. Furthermore, we show how our approach enables any physics-based load model to be incorporated into the power flow formulation directly, without requiring simplifying assumptions. Importantly, the use of I-V formulation and the modeling formalism that it entails can unify power system analyses [9] such that consistent results are obtained between transient and stead-state analyses. Additionally, it has been postulated that the use of voltage and current state variable may also enable superior optimal power flow formulations [10].

Our proposed simulation approach is based on using the equivalent circuit formulation presented in [11]-[13]. With this equivalent circuit framework, we are able to apply techniques from circuit simulation, such as variable limiting and continuation methods, to robustly solve the general power flow problem. We apply our approach to systems as large as the 9241 bus test system and demonstrate convergence to the correct physical solution for all cases, independent of the initial guesses for the unspecified Q variables.

## II. BACKGROUND

### A. Equivalent Circuit Formulation

The equivalent circuit approach for generalized modeling of the steady-state power system response (i.e. power flow and three-phase power flow) was recently introduced in [11]-[13]. This circuit-based formulation represents both the transmission and distribution power grid as an aggregation of circuit elements. Each of the power system components (including constant power models, i.e. PQ and PV buses) is represented by an equivalent circuit model based on the underlying current and voltage state variables without loss of generality. This formulation can represent any physics based load model or measurement based semi-empirical load model as a sub-circuit that can then be combined hierarchically with other circuit abstractions to build larger aggregated models. The equivalent circuit representations of the most prominent models for the power flow problem are described in the following sections.

#### 1) PV Bus

The equivalent circuit formulation provides a choice to model the constant voltage (PV) node as either a complex voltage source (in terms of complex current) [11] or a complex current source (in terms of complex voltage) [13]. It has been shown that representing the PV bus as a complex current source offers superior convergence [13] for Newton-Raphson (NR) iterations. To enable the application of NR, the complex current source is split into real and imaginary current sources ($I_{RG}$ and $I_{IG}$, respectively). This is necessary due to the non-analyticity of complex conjugate function [13]:

$$I_{RG} = \frac{P_G V_{RG} + Q_G V_{IG}}{V_{RG}^2 + V_{IG}^2} \tag{1}$$

$$I_{IG} = \frac{P_G V_{IG} - Q_G V_{RG}}{V_{RG}^2 + V_{IG}^2} \tag{2}$$

An additional constraint equation for the voltage magnitude is needed for the PV model to ensure that the voltage is equal to its set point but also to compensate for the fact that there is an additional unknown variable for the reactive power $Q_G$, i.e.

$$V_G = V_{RG}^2 + V_{IG}^2 \tag{3}$$

The first two terms of the Taylor expansions for equations (1) through (3) can be used to linearize the functions and derive an equivalent circuit model, as shown in Fig. 1. For example, linearization of the real generator current is:

$$I_{RG}^{k+1} = \frac{\partial I_{RG}}{\partial Q_G}\Big|_{Q_G^k, V_{RG}^k, V_{IG}^k}(Q_G^{k+1}) + \frac{\partial I_{RG}}{\partial V_{RG}}\Big|_{Q_G^k, V_{RG}^k, V_{IG}^k}(V_{RG}^{k+1}) \\
+ \frac{\partial I_{RG}}{\partial V_{IG}}\Big|_{Q_G^k, V_{RG}^k, V_{IG}^k}(V_{IG}^{k+1}) + I_{RG}^k - \frac{\partial I_{RG}}{\partial Q_G}\Big|_{Q_G^k, V_{RG}^k, V_{IG}^k}(Q_G^k) \\
- \frac{\partial I_{RG}}{\partial V_{RG}}\Big|_{Q_G^k, V_{RG}^k, V_{IG}^k}(V_{RG}^k) - \frac{\partial I_{RG}}{\partial V_{IG}}\Big|_{Q_G^k, V_{RG}^k, V_{IG}^k}(V_{IG}^k) \tag{4}$$

The first term in (4) represents a current source that is a function of the reactive power; the second term represents a conductance, since the real current is proportional to the real voltage; the third term represents a voltage-controlled current source, since the real current is proportional to the imaginary voltage. The remaining terms are all dependent on known values from the previous iteration, so they can be lumped together and represented as an independent current source.

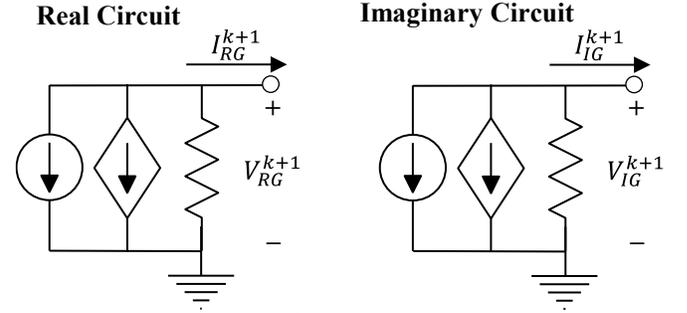

*Figure 1: Equivalent Circuit Model for PV generator model.*

#### 2) PQ Bus

Similar to the PV bus, the constant power node (PQ bus) can also be represented as an equivalent circuit via either a complex voltage source or a complex current source. It has been empirically determined that superior convergence is observed when the load bus is modeled as complex current source. The two fundamental equations for the real and imaginary currents for the PQ buses are given by:

$$I_{RL} = \frac{P_L V_{RL} + Q_L V_{IL}}{V_{RL}^2 + V_{IL}^2} \tag{5}$$

$$I_{IL} = \frac{P_L V_{IL} - Q_L V_{RL}}{V_{RL}^2 + V_{IL}^2} \tag{6}$$

Linearizing the load model in (5)-(6) via Taylor expansion results in three elements in parallel in both circuits: a conductance, a voltage-controlled current source, and an independent current source.

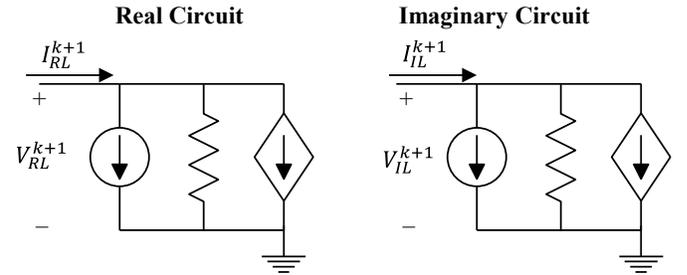

*Figure 2: Equivalent split-circuit PQ load model.*

#### 3) Physics Based Load Model

It has been previously shown in [9] that any physics based device model can also be directly incorporated into the equivalent circuit formulation. For instance, consider the three-phase induction motor (IM) example that was previously discussed in detail in [9]. The steady state and transient behavior of an IM can be expressed by a set of five ordinary differential equations. These mathematical expressions can be mapped into an equivalent circuit as shown in Fig. 3 with the use of standard circuit simulation techniques [14]. Due to the use of DQ transformation [15], this physics based equivalent circuit model of an IM can be directly used for steady state power flow formulations by shorting the inductors and open circuiting the capacitors.

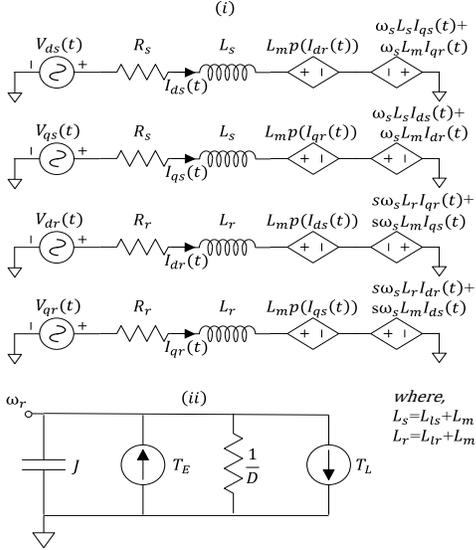

Figure 3: Equivalent circuit for the three-phase induction motor model in natural state variables of I-V.

*4) Measurement Based Load and Generation Models*

Aggregated load and generation in the grid can also be represented as an equivalent circuit based on our semi-empirical approach in [8]. It was shown that finite order Taylor expansion with use of complex rectangular voltage and current state variables can be used to model any power grid component that is described by such variables in terms of measurement data from the grid.

The system load and generation when modeled as current injections is given by (7) and is represented by a split equivalent circuit shown in Fig. 2. Similarly, system load and generation when modeled as voltage sources is given by (8) and is represented by a split equivalent circuit in Fig. 4:

$$I_C = g_1^C + g_2^C V_R + g_3^C V_I + g_4^C V_R V_I + g_5^C V_R^2 + g_6^C V_I^2 \ldots \quad (7)$$

$$V_C = g_1^C + g_2^C I_R + g_3^C I_I + g_4^C I_R I_I + g_5^C I_R^2 + g_6^C I_I^2 \ldots \quad (8)$$

where, $C \in \{R, I\}$ represents the placeholder for real and imaginary parts and $g_i^C$ represents the optimal coefficients for the semi-empirical model.

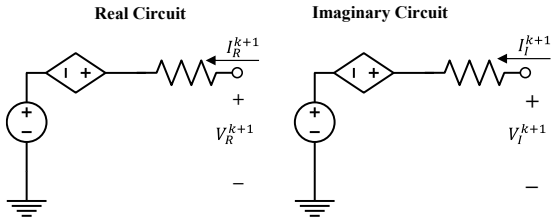

Figure 4: Equivalent split circuit for semi-empirical model based on current dependent variables.

*B. Tree Link Analysis*

To formulate the circuit equations using the equivalent circuit components, a graph-theoretic tree-link (TLA) method [16] is applied to formulate the equivalent circuit equations of the linearized equivalent circuit. TLA has been shown to perform seamlessly for balanced power flow, three-phase power flow, transient and harmonic power flow analyses in [11]-[13],[9] and [17], offering superior conditioning of the matrix equations over existing nodal methods (MNA). MNA is generally used for circuit simulations of electronic systems due to its simplicity and efficiency. However, TLA is known to provide superior numerical conditioning and the ability to accommodate both voltage and current state variables inherently.

For applications in three-phase power system analyses, the ability of TLA to naturally incorporate current state variables enables the handling of a large number of coupled inductors typical to power grid models. Furthermore, the TLA formulation is capable of accommodating ideal switches (switching from zero impedance to zero conductance), which enables a straightforward inclusion of components that are switched into and out of the grid into the power flow simulation. This capability is particularly helpful for contingency analyses.

### III. CIRCUIT SIMULATION TECHNIQUES

Decades of research in circuit simulation has demonstrated that continuation methods and homotopy can be applied for determining the DC state of a circuit using NR. These techniques have been shown to make NR robust and practical for large-scale circuit problems [18]. Most notable is the ability to guarantee convergence to the correct physical solution (i.e. global convergence) and the capability of finding multiple operating points [18]. In this paper, we propose analogous techniques for ensuring convergence to the correct physical solution for the power flow and three-phase power flow problems.

*A. Variable Limiting*

The solution space of the system node voltages in a power flow problem is well defined. While solving the power flow problem, a large NR step may step out of this solution space and result in either non-convergence or convergence to a non-physical solution. It is therefore important to limit the NR step before it makes an invalid step out of the solution space. We propose to apply variable limiting to achieve the postulated goal. In this technique, the state variables that are most sensitive to initial guesses are damped when the NR algorithm takes a large step out of the pre-defined solution space. Note that not all of the variables are damped in variable limiting, and circuit simulation research has shown that it provides superior convergence compared to damped NR in general.

In the power flow problem, the voltages on the PV node are highly sensitive to the reactive power (Q) value at that node. In the I-V formulation of the power flow problem each PV node augments the solution space by an additional unknown variable Q for which an initial guess has to be assigned. However, unlike the node voltages, it is very hard to choose the appropriate initial guess for these Q variables as they exhibit a large solution space. Therefore, with an arbitrary choice of initial values, the power flow problem may diverge or converge to the wrong solution.

In order to tackle this problem, the voltages at the PV node are damped during the NR iterations whenever they make a large step out of the pre-defined solution space. Fig. 5 can be used to demonstrate this graphically. The plot in Fig. 5 shows results for a 2383 bus test system that was represented in I-V formulation and simulations were run on it for six different initial guesses for unspecified Q. The maximum bus voltage from the solution of the power flow problem for each initial guess was then plotted for two scenarios: without and with variable limiting enabled. The plots in the figure show that when variable limiting is not enabled, the voltage solution diverges to very high magnitudes (up to $10^4$) and may not converge even in 100 iterations. However, when the variable limiting option is enabled, divergence is not observed and the bounded bus voltages result in fast convergence.

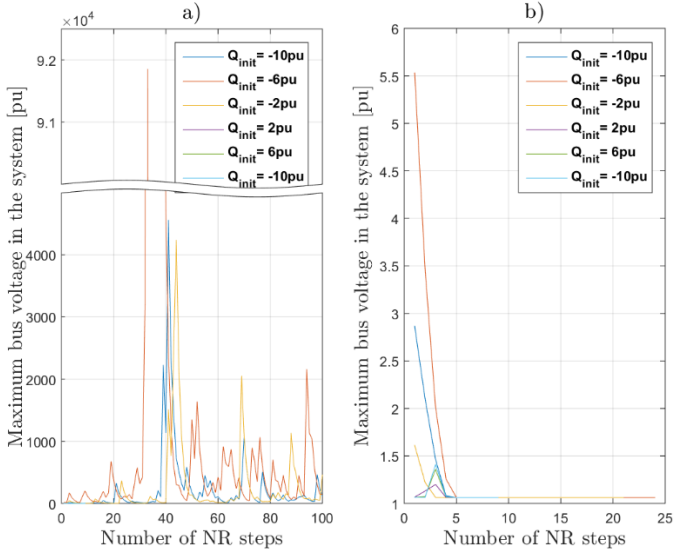

Figure 5: Voltage profile for maximum bus voltage in 2383 Bus System: a) w/o Variable Limiting b) with Variable Limiting

In order to apply variable limiting in our prototype simulator, the mathematical expressions for the PV nodes in the system are modified as follows:

$$I_{CG}^{k+1} = \alpha \frac{\partial I_{CG}}{\partial V_{RG}}|_{Q_G^k, V_{RG}^k, V_{IG}^k}(V_{RG}^{k+1} - V_{RG}^k) + I_{CG}^k \quad (9)$$
$$+ \alpha \frac{\partial I_{CG}}{\partial V_{IG}}|_{Q_G^k, V_{RG}^k, V_{IG}^k}(V_{IG}^{k+1} - V_{IG}^k)$$
$$+ \frac{\partial I_{CG}}{\partial Q_G}|_{Q_G^k, V_{RG}^k, V_{IG}^k}(Q_G^{k+1} - Q_G^k)$$

where, $0 \leq \alpha \leq 1$ and $C \in \{R, I\}$ represents the placeholder for real and imaginary parts. The magnitude of $\alpha$ is dynamically varied through heuristics such that convergence to correct physical solution is achieved in the most efficient manner.

### B. Dynamic Power Stepping

As with large circuit simulation problems, variable limiting alone is insufficient for solving some of the most complex large-scale power flow problems. Variable limiting alone fails for cases where the Jacobian matrix is close to being singular or when the solution matrix remains out of the expected bounds of the solution space over multiple iterations. For such cases, we find it necessary to apply a continuation method, such as the power stepping described in [11], which is analogous to the source stepping and gmin stepping approaches in standard circuit simulation solvers.

The power stepping technique is a continuation method approach wherein the magnitudes of loads and generations in the system are scaled back by a factor of $\beta$. If these loads and generations are scaled down all the way to zero, then the PQ buses in the system are expressed by purely linear equations. Similarly, the real power dependent current source non-linearities of the PV buses are also eliminated. Therefore, by applying the power stepping factor, the non-linearities in the system are greatly eased and convergence is easily achieved. Upon convergence, the factor is gradually scaled back up to unity in order to solve the original problem. In this method, as in all continuation methods, the solution from the prior step is used as the initial condition for the next step:

$$\forall i \in PV: P_i = \beta P_i \quad (10)$$
$$\forall i \in PQ: P_i = \beta P_i \text{ and } Q_i = \beta Q_i \quad (11)$$

where, *PQ* are all PQ buses and *PV* are all PV buses

It is worth noting that this technique does not improve the convergence properties of traditional 'PQV' power flow formulation. This is because scaling P and Q does not affect the Jacobian.

### IV. RESULTS

In this section, example cases are simulated in our prototype solver SUGAR (Simulation with Unified Grid Analyses and Renewables) to validate the superior performance offered by our equivalent circuit formulation approach. The example cases affirm that the proposed framework can guarantee convergence to correct physical solutions for all power flow cases, independent of the choice of the initial guess.

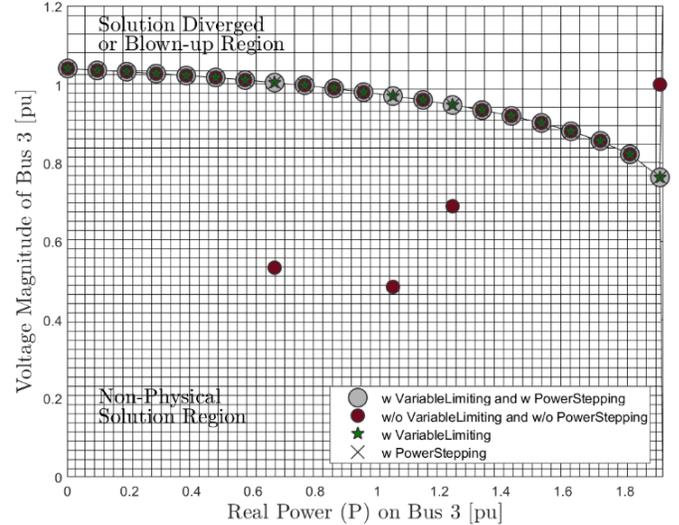

Figure 6: Solution of Bus 3 voltage for IEEE 14 bus test system with increasing loading factors with and without circuit simulation methods

In the first example, simulations are run on IEEE 14 bus test system (from flat start) in steps of increasing loading factors (up to 4x) for the following four scenarios: 1) both power stepping and variable limiting option disabled, 2) with power stepping option enabled and variable limiting disabled, 3) with variable limiting option enabled and power stepping disabled, and 4) both power stepping and variable limiting option enabled. The solutions for the bus 3 voltage magnitude at the end of each simulation are then plotted in Fig. 6. The plot shows that convergence to the correct physical solution is achieved for each simulation instance when either variable limiting or power stepping option is enabled. However, without these options enabled in SUGAR, the solution in many simulation instances has either converged to the wrong solution or diverged altogether.

It is important to note that the use of power stepping or variable limiting alone may not guarantee convergence to the correct physical solution as the complexity and the size of the system increases. This is demonstrated in the second example wherein power flow simulations are run on large 2869 and 9241 bus test systems.

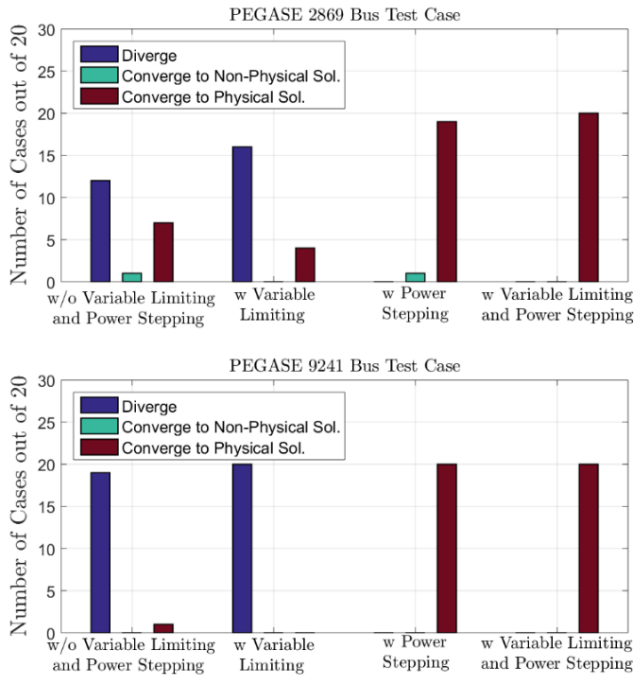

*Figure 7: Power flow results for 2869 bus and 9241 bus test systems with and without circuit simulation techniques*

In this example, power flow simulations are run on both the 2869 and 9241 bus test systems for 20 different initial guesses for Q values that are uniformly distributed in the range of -10 pu and 10 pu. All 20 simulations are run for each of these solver settings under the same four scenarios. The convergence results plotted in Fig. 7 show that without circuit simulation techniques, most of the cases in these systems either diverge or converge to the wrong solution. Convergence to the correct physical solution is only observed when both variable limiting and power stepping are enabled. In general, the two techniques discussed in this paper are by no means an exhaustive list of circuit simulation techniques that can be applied to the power flow problem or the three-phase power flow problem. Other circuit simulation techniques can also be incorporated into the equivalent circuit framework for even more robust and efficient simulations.

## V. CONCLUSIONS

In this paper, the power flow problem is formulated using an equivalent circuit framework that when combined with robust circuit simulation techniques can guarantee convergence to the correct physical solution. This work directly addresses known convergence issues presented by existing methods that use current-voltage (I-V) formulation for power flow analysis.